\documentclass[11pt]{article}

\usepackage[final]{acl}

\usepackage{times}
\usepackage{latexsym}
\usepackage[T1]{fontenc}
\usepackage[utf8]{inputenc}
\usepackage{microtype}
\usepackage{graphicx}
\usepackage{caption}
\usepackage{subcaption}
\usepackage{tcolorbox}

\usepackage{amsmath,amssymb,amsthm}
\usepackage{algorithm}
\usepackage{algorithmic}

\usepackage{booktabs}
\usepackage{multirow}
\usepackage{makecell}
\usepackage{xcolor}
\usepackage{url}
\usepackage{enumitem}
\usepackage[normalem]{ulem}

\newtheorem{definition}{Definition}

\newcommand{\eps}{\varepsilon}
\newcommand{\R}{\mathbb{R}}

\newcommand{\Prob}{\mathbb{P}}
\newcommand{\calM}{\mathcal{M}}
\newcommand{\calD}{\mathcal{D}}
\newcommand{\calP}{\mathcal{P}}
\newcommand{\calO}{\mathcal{O}}

\title{DP-ES: Differentially Private Evolution Strategies for Prompt Optimization}

\author{
\textbf{Ziniu Liu}\textsuperscript{1},
\textbf{Aiping Li}\textsuperscript{1,*},
\textbf{Yue Han}\textsuperscript{1},
\textbf{Han Yu}\textsuperscript{1},\\
\textbf{Junjian Zhang}\textsuperscript{1},
\textbf{Dong Zhu}\textsuperscript{1},
\textbf{Changjian Li}\textsuperscript{1},
\textbf{Shiqiang Zhang}\textsuperscript{2}\\
\textsuperscript{1}National University of Defense Technology\\
\textsuperscript{2}CRRC Zhuzhou Electric Locomotive Research Institute Co., Ltd.,\\
China Academy of Railway Sciences\\
{\small\textsuperscript{1}\texttt{\{liuzn\_nudt,liaiping,hanyue,yuhan17,zhangjunjian24,zhud,lichangjian23\}@nudt.edu.cn}}\\[-0.2em]
{\small\textsuperscript{2}\href{mailto:zhangsq5@csrzic.com}{zhangsq5@csrzic.com}\quad
\textsuperscript{*}Corresponding author: \href{mailto:liaiping@nudt.edu.cn}{liaiping@nudt.edu.cn}}
}

\begin{document}
\maketitle

\begin{abstract}
Token-level differentially private (DP) prompt optimization methods such as DP-OPT can become unstable under tight privacy budgets: on GSM8K, DP-OPT obtains $49.5{\pm}28.5\%$ across 30 runs, and a logged search trajectory reveals prompt-template drift and noise-sensitive irreversible choices. We diagnose these as structural consequences of greedy token-by-token construction over privately aggregated counts. We then propose \textbf{DP-ES} (Differentially Private Evolution Strategies), a structurally cleaner alternative that maintains a population of full prompts, mutates them via LLM calls that never access the private dataset, and spends privacy only on sampled-Gaussian evaluation; deterministic or Gumbel-smoothed selection is post-processing. Under a conservative $(\varepsilon{\leq}1.0, \delta{=}10^{-5})$ guarantee, DP-ES achieves \textbf{88.1\% on GSM8K} (+38.6 pp over DP-OPT, $\sim$9$\times$ lower std), \textbf{99.7\% on MedQA}, \textbf{73.5\% on BANKING77}, and \textbf{86.8\% on Alpaca}. It is also 2.5$\times$ faster in wall-clock time and uses 3.3$\times$ fewer logged private-data call groups than DP-OPT. Selection and population ablations, implementation-level noise checks, and a 200-profile exact-match memorization stress test complement the formal guarantee. \textbf{Scope.} Our experiments establish optimization robustness under DP noise, especially where prompt structure is critical; end-to-end validation on genuinely sensitive, non-saturated deployment data remains future work. Code: \href{https://github.com/StephCpa/dp-es}{github.com/StephCpa/dp-es}.
\end{abstract}

\section{Introduction}
\label{sec:introduction}

Large Language Models (LLMs) have demonstrated remarkable capabilities through in-context learning \cite{brown2020language,ouyang2022training}, yet their performance heavily relies on carefully crafted prompts \cite{liu2023pre}. Automatic prompt optimization \cite{yuksekgonul2024textgrad,yang2023large} can bridge this gap, but when the optimization data are sensitive (e.g., patient records, financial logs), the optimizer can memorize and expose private information, as evidenced by recent membership inference and extraction attacks \cite{wang2024decodingtrust,rahman2024sok,fu2023practical}. For instance, a naive optimizer might generate a prompt like \textit{``Remember how patient \#12847 had atrial fibrillation...''}, directly violating privacy. Any optimizer intended for such settings must therefore satisfy formal \textbf{Differential Privacy (DP)} guarantees \cite{dwork2006calibrating}---and, crucially, the optimization mechanism itself must remain robust under the noise that DP introduces, which is the question we focus on in this work.

\noindent\textbf{The Challenge: Task Complexity and Privacy Budgets.}
Applying DP to prompt optimization is non-trivial due to the discrete nature of text. Recent work like DP-OPT \cite{hong2024dpopt} uses histogram-based token generation with the exponential mechanism. Other approaches apply DP to in-context learning via private few-shot generation or nearest neighbor retrieval \cite{tang2024privacy,koskela2025differentially}, but they rely on additional retrieval or embedding machinery. While DP-OPT is effective on simple tasks, it is unstable on our complex-reasoning setting: across 30 GSM8K runs it obtains $49.5{\pm}28.5\%$, and trajectory analysis reveals prompt-template drift and noise-sensitive irreversible decisions (Appendix~\ref{sec:appendix:dpopt_failures}). This fragility is consistent with token-level myopia and non-backtracking search under tight privacy constraints.

\noindent\textbf{Private Evolution Perspective.}
Private Evolution (PE) is a training-free DP paradigm for data synthesis via APIs \cite{lin2024dpsda_images,xie2024dpsda_text,lin2025simpe}. Its core advantage is structural: only low-sensitivity evaluation touches private data, while exploration/generation and selection over privatized scores are post-processing. This yields a decoupling between exploration and privacy expenditure---increasing the number or strength of mutations improves search without consuming additional privacy budget. DP-ES is a prompt-optimization instantiation of this PE principle.

\begin{figure*}[t]
\centering
\includegraphics[width=\textwidth]{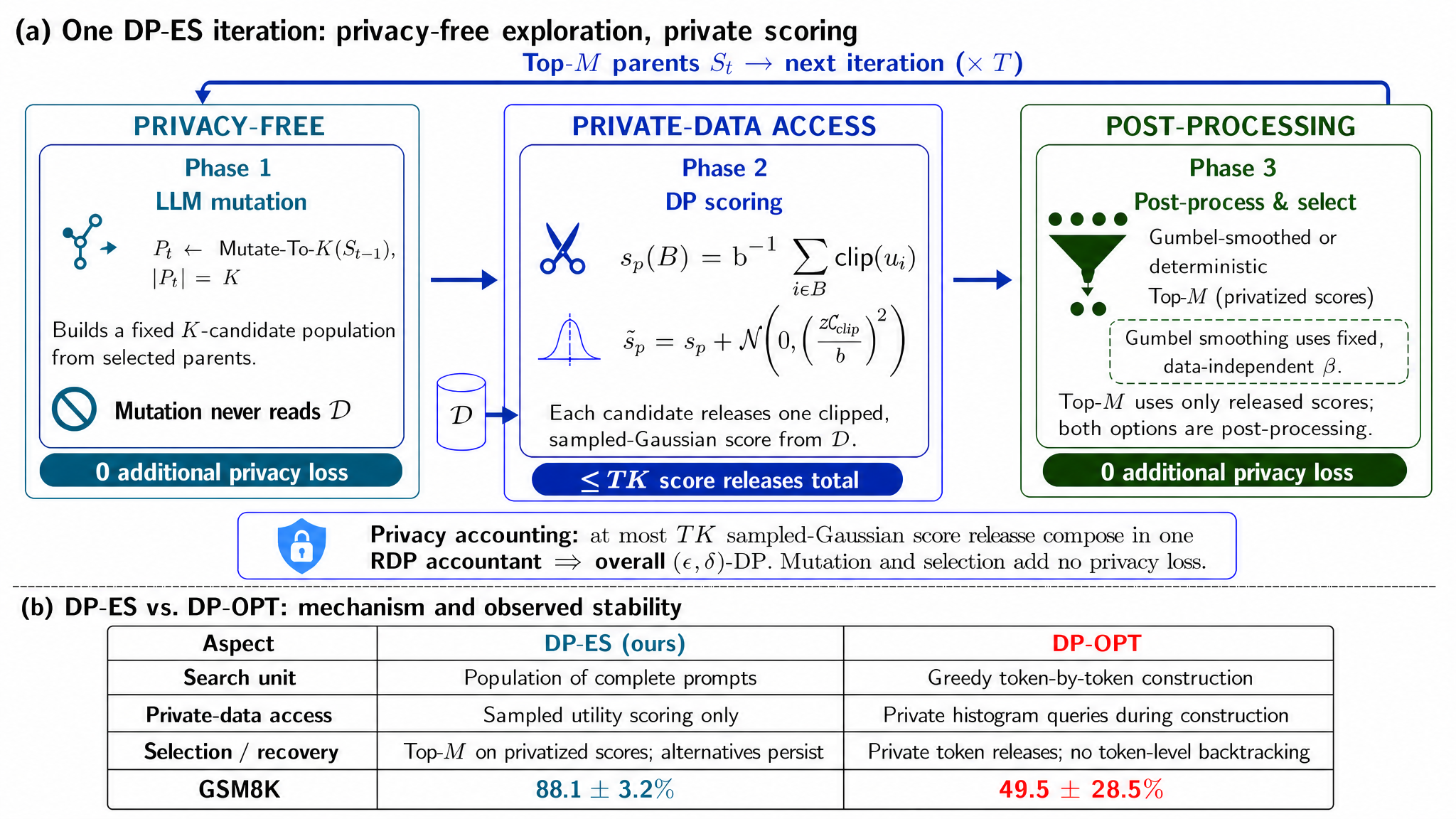}
\caption{\textbf{DP-ES workflow and mechanism-level contrast with DP-OPT.} Only sampled-Gaussian scoring accesses the private dataset. LLM mutation and deterministic or Gumbel-smoothed top-$M$ parent selection operate without further private-data access and add no privacy loss. Mechanism rows summarize canonical DP-OPT; the bottom panel reports our black-box adaptation's GSM8K mean and standard deviation across 30 runs without imposing a post-hoc failure threshold.}
\label{fig:overview}
\end{figure*}

\noindent\textbf{Our Contribution: DP-ES.}
We frame this work as a \emph{diagnostic and methodological} contribution for DP prompt optimization: we systematically characterize the instability mechanisms of token-level DP construction under tight privacy budgets and propose \textbf{Differentially Private Evolution Strategies (DP-ES)}, a structurally cleaner alternative. Unlike methods that build prompts token by token from privately aggregated counts, DP-ES maintains a population of full prompts that evolve over time. The key design choice is to decouple exploration from privacy consumption: using LLMs to generate semantic mutations consumes \textbf{zero privacy budget} (by the post-processing property), allowing the scarce privacy budget to be allocated strictly to sampled-Gaussian evaluation.

Our contributions are:
\vspace{-0.3em}
\begin{enumerate}[leftmargin=*,itemsep=-0.3em]
    \item \textbf{Instability diagnosis:} A 30-run analysis of DP-OPT on GSM8K, complemented by a concrete logged trajectory that exposes prompt-template drift and noise-sensitive irreversible selection in token-level construction (Section~\ref{sec:results}).
    \item \textbf{Method:} A structurally cleaner evolution-based DP prompt optimizer with privacy-free mutations, sampled-Gaussian scoring under RDP accounting, and deterministic or Gumbel-smoothed post-processing selection.
    \item \textbf{Empirical robustness:} $\sim$9$\times$ lower standard deviation than DP-OPT on GSM8K, plus strong or competitive results on MedQA, BANKING77, and Alpaca under a conservative $(\varepsilon{\leq}1,\delta{=}10^{-5})$ guarantee.
    \item \textbf{Efficiency:} 2.5$\times$ faster wall-clock runtime and 3.3$\times$ fewer logged private-data call groups than DP-OPT on GSM8K, with an explicit generation-vs-evaluation cost decomposition.
    \item \textbf{Privacy evidence:} A reproducible RDP accounting calculation and noise-distribution checks, plus a 200-profile synthetic exact-match stress test (0 PII strings found for DP-ES vs.\ 40 for an adversarial non-DP control).
\end{enumerate}

\noindent\textbf{Scope of Empirical Evidence.}
Our experiments establish DP-ES as a \emph{structurally more robust DP prompt optimizer} than token-level alternatives, especially on tasks where prompt structure is critical. They do not by themselves validate deployment on genuinely sensitive, non-saturated datasets. The formal $(\varepsilon,\delta)$-DP guarantee bounds record-level leakage independently of the benchmark, while the implementation checks and exact-match stress test provide complementary diagnostics. We make these scope boundaries explicit in the Limitations section.

\section{Related Work}
\label{sec:related}

\noindent\textbf{Prompt Optimization.}
Gradient-free methods such as APE \cite{zhou2023large} and OPRO \cite{yang2023large} search over discrete prompt candidates using LLM-generated variations, while gradient-based approaches like TextGrad \cite{yuksekgonul2024textgrad} use LLM feedback as differentiable signals. Both families optimize prompts effectively but assume unrestricted access to training data with no privacy guarantees \cite{edemacu2024privacy}, making them unsuitable for sensitive domains such as healthcare, finance, and education where data protection is legally mandated.

\noindent\textbf{Differential Privacy in Machine Learning.}
DP-SGD \cite{abadi2016deep} and PATE \cite{papernot2018scalable} provide formal $(\varepsilon,\delta)$-DP guarantees for continuous model parameters during training. DP-LoRA \cite{liu2024differentially,frery2025private} extends this to parameter-efficient fine-tuning of LLMs, achieving strong performance with reduced memory overhead. However, all these methods require access to model weights, precluding their use with black-box API-only LLMs---the dominant deployment mode for state-of-the-art models such as GPT-4 and DeepSeek.

\noindent\textbf{DP Prompt Optimization.}
For discrete prompts over black-box LLMs, DP-OPT \cite{hong2024dpopt} uses histogram-based token generation with the exponential mechanism, constructing prompts token by token from privately aggregated counts. While effective on classification and factual QA, this greedy token-level approach is unstable in our GSM8K experiments and can exhibit prompt-template drift and noise-sensitive irreversible decisions (Section~\ref{sec:results}).

\noindent\textbf{Other Approaches.}
Other directions include DP few-shot demonstration generation for in-context learning \cite{tang2024privacy}, which privately selects exemplars but cannot optimize the instruction prompt itself; private retrieval augmentation \cite{koskela2025differentially}, which protects retrieval queries but inherits the limitations of fixed prompt templates; and PromptDPSGD \cite{Duan2023FlocksOS}, which applies DP-SGD to continuous soft prompt embeddings but requires white-box model access and struggles on smaller models (Appendix~\ref{sec:appendix:local}).

\noindent\textbf{Private Evolution.}
Private Evolution (PE) \cite{lin2024dpsda_images,xie2024dpsda_text,lin2025simpe} is a training-free DP framework originally developed for data synthesis via APIs. Its key structural advantage is that only a low-sensitivity evaluation step touches private data, while API-based generation is pure post-processing. This creates a favorable decoupling: exploration can be arbitrarily rich without consuming privacy budget. DP-ES adapts this principle from data synthesis to prompt optimization, inheriting the favorable privacy-exploration trade-off while introducing population-based search over the discrete prompt space.

\noindent\textbf{Evolution Strategies and DP.}
Evolution strategies (ES) \cite{rechenberg1973evolutionsstrategie} are derivative-free algorithms that maintain and evolve populations of candidate solutions. Prior non-private prompt optimizers, including EvoPrompt \cite{guo2024evoprompt} and PromptBreeder \cite{fernando2023promptbreeder}, already use LLM-driven mutation and population-based selection to optimize discrete prompts. DP-ES builds on this evolutionary-prompt-optimization line while introducing record-level DP through sampled-Gaussian scoring and post-processing selection. The population-based design provides natural robustness to noise-induced regressions---a critical advantage under tight privacy budgets where evaluation noise is substantial.

\noindent\textbf{Privacy Auditing.}
Empirical privacy attacks and audits can reveal implementation errors but cannot certify DP from a finite number of trials \cite{nasr2021adversary,jagielski2020auditing}. We therefore base our guarantee on mechanism-level RDP accounting and use noise checks and an exact-match memorization stress test only as complementary diagnostics (Section~\ref{sec:privacy_analysis}).

\noindent\textbf{Positioning of DP-ES.}
Table~\ref{tab:comparison} summarizes the landscape. DP-ES uniquely combines evolutionary exploration with sampled-Gaussian scoring and post-processing selection, filling the gap for a robust, black-box-compatible DP prompt optimization method that scales across diverse task complexities---from structured classification to multi-step reasoning.

\begin{table}[t]
\centering
\small
\setlength{\tabcolsep}{3pt}
\begin{tabular}{@{}lccc@{}}
\toprule
\textbf{Method} & \textbf{Priv.} & \textbf{Disc.} & \textbf{Mechanism} \\
\midrule
TextGrad & \texttimes & \checkmark & LLM gradients \\
OPRO & \texttimes & \checkmark & LLM scoring \\
EvoPrompt & \texttimes & \checkmark & Evolutionary search \\
PromptBreeder & \texttimes & \checkmark & Evolutionary search \\
\midrule
DP-SGD & \checkmark & \texttimes & Gaussian noise \\
DP-OPT & \checkmark & \checkmark & Histogram + Exp \\
\midrule
\textbf{DP-ES} & \checkmark & \checkmark & \textbf{Evol. + Gauss} \\
\bottomrule
\end{tabular}
\caption{Comparison of prompt optimization methods. DP-ES adds formal record-level DP to evolutionary optimization over discrete prompts.}
\label{tab:comparison}
\end{table}
\vspace{-1em}

\vspace{0.5em}
\section{Methodology: DP-ES}
\label{sec:methodology}

\noindent\textbf{Problem Formulation.}
Let $\calD = \{(x_i, y_i)\}_{i=1}^n$ be a private dataset where $x_i$ represents task inputs and $y_i$ represents desired outputs (e.g., question-answer pairs). Let $\calP$ denote the space of natural language prompts. Our goal is to find a prompt $p^* \in \calP$ that maximizes task performance:
\begin{equation}
p^* = \arg\max_{p \in \calP} f(p, \calD)
\end{equation}
where $f: \calP \times \calD \to \R$ measures prompt quality (e.g., accuracy on $\calD$).

\noindent\textbf{Privacy requirement:} The optimization algorithm must satisfy $(\eps, \delta)$-differential privacy with respect to $\calD$, ensuring that any single record $(x_i, y_i)$ has bounded influence on the output prompt.
Table~\ref{tab:notation} summarizes the notation used below.

\begin{table}[t]
\centering
\scriptsize
\setlength{\tabcolsep}{3pt}
\begin{tabular}{cl@{\qquad}cl}
\toprule
$\calD,n$ & private data, size & $K,M,T$ & candidates, parents, rounds \\
$b,q$ & batch size, $q=b/n$ & $z$ & Gaussian noise multiplier \\
$u_i(p)$ & per-record utility & $C_{\text{clip}}$ & clipping bound \\
$\widetilde{s}_p$ & privatized score & $\beta$ & Gumbel smoothing scale \\
\bottomrule
\end{tabular}
\caption{Notation used in DP-ES.}
\label{tab:notation}
\end{table}

\vspace{-0.5em}
\begin{definition}[$(\eps, \delta)$-Differential Privacy]
\label{def:dp}
A randomized algorithm $\calM: \calD \to \calO$ satisfies $(\eps, \delta)$-DP if for all neighboring datasets $\calD \sim \calD'$ (differing in one record) and all measurable sets $S \subseteq \text{Range}(\calM)$:
\vspace{-0.8em}
\begin{equation}
\Prob[\calM(\calD) \in S] \leq e^\eps \cdot \Prob[\calM(\calD') \in S] + \delta
\end{equation}
\end{definition}

\vspace{-0.5em}
\noindent\textbf{DP-ES Algorithm Overview.}
DP-ES (Algorithm~\ref{alg:dpes}) employs an evolution strategy with three key components:

\begin{enumerate}
    \vspace{-0.8em}
    \item \textbf{Mutation:} Generate prompt variations using an LLM (privacy-free)
    \vspace{-0.8em}
    \item \textbf{DP Evaluation:} Score prompts on $\calD$ with calibrated noise
     \vspace{-0.8em}
    \item \textbf{Selection:} Select top prompts from privatized scores (Gumbel-smoothed or deterministic)
\end{enumerate}

\begin{nolinenumbers}
\begin{algorithm}[t]
\caption{DP-ES: Differentially Private Evolution Strategies}
\label{alg:dpes}
\small
\begin{algorithmic}[1]
\REQUIRE Private dataset $\calD$, privacy budget $(\eps_{\text{total}}, \delta)$, initial prompt $p_0$, population size $K$, parent count $M\leq K$, iterations $T$
\ENSURE Optimized prompt $p^*$ with $(\eps_{\text{total}}, \delta)$-DP
\STATE Initialize parent set $\mathcal{S}_0 \leftarrow \{p_0\}$ and best-so-far score $\tilde{s}^{*}\leftarrow-\infty$
\STATE Calibrate $z$ for at most $T K$ sampled-Gaussian scores \COMMENT{RDP accountant}
\FOR{$t = 1$ to $T$}
    \STATE \textcolor{blue}{// Phase 1: Privacy-free mutation}
    \STATE $\mathcal{P}_t \leftarrow \text{Mutate-To-}K(\mathcal{S}_{t-1})$ \COMMENT{$|\mathcal{P}_t|=K$}

    \STATE \textcolor{blue}{// Phase 2: Sampled-Gaussian scoring (RDP-accounted)}
    \FOR{each $p \in \mathcal{P}_t$}
        \STATE Sample $B\subset\calD$, $|B|=b$; set $u_i(p)=\mathbb{1}[\text{LLM}(p,x_i)=y_i]$
        \STATE $s_p = \frac{1}{b}\sum_{i\in B}\text{clip}(u_i(p),[0,C_{\text{clip}}])$ \COMMENT{$C_{\text{clip}}=1.0$}
        \STATE Add noise: $\tilde{s}_p = s_p + \mathcal{N}(0, \sigma^2)$ \COMMENT{$\sigma=zC_{\text{clip}}/b$}
    \ENDFOR
    \STATE Update $p^*,\tilde{s}^*$ using $\arg\max_{p\in\mathcal{P}_t}\tilde{s}_p$ \COMMENT{post-processing}

    \STATE \textcolor{blue}{// Phase 3: Selection over privatized scores}
    \STATE \textbf{Option A} (default): $\hat{s}_p \leftarrow \tilde{s}_p + \text{Gumbel}(0,\beta)$  \COMMENT{fixed $\beta$; post-processing}
    \STATE \textbf{Option B}: $\hat{s}_p \leftarrow \tilde{s}_p$ \COMMENT{Deterministic top-$M$; pure post-processing}
    \STATE $\mathcal{S}_t \leftarrow \text{top-}M(\mathcal{P}_t, \{\hat{s}_p\})$ \COMMENT{parents for next round}
\ENDFOR
\RETURN $p^*$
\end{algorithmic}
\end{algorithm}
\end{nolinenumbers}

\noindent\textbf{Privacy-Free Mutations.}
\textbf{Key Innovation:} Unlike gradient-based methods that require DP protection for gradient computation, DP-ES sends only a public or previously privatized parent prompt to the mutation LLM; neither records from $\calD$ nor record-derived feedback are included. Mutation is therefore post-processing and consumes zero additional privacy budget.

\refstepcounter{theorem}\label{prop:mutation-free}
\noindent\textbf{Proposition~\thetheorem\ (Mutation Privacy-Freeness).}
\textit{Let $\text{Mutate}: \calP \to \calP$ be a function that generates a new prompt from a parent prompt using an LLM, without accessing the private dataset $\calD$. Then Mutate consumes zero privacy budget.}

\noindent\textbf{Proof.} See Appendix~\ref*{sec:appendix:proofs}.

\refstepcounter{theorem}\label{prop:decoupled}
\noindent\textbf{Proposition~\thetheorem\ (Decoupled Exploration).}
\textit{If mutation does not access $\calD$, then the privacy loss of DP-ES depends only on the sampled-Gaussian score releases and is independent of the number or strength of mutations per iteration.}
\noindent\textbf{Proof.} See Appendix~\ref*{sec:appendix:proofs}.

\noindent \textbf{Mutation strategies:} We implement 3 modes (EXPLORE, EXPLOIT, BALANCED); details and prompt templates are provided in Appendix~\ref*{sec:appendix:exp}.

\noindent\textbf{DP Evaluation via Gaussian Mechanism.}
After mutation, each candidate is evaluated on a uniformly sampled batch $B\subset\calD$ of size $b$. Its score is the mean of clipped per-record utilities:

\vspace{-1em}
\begin{equation}
\begin{aligned}
u_i(p) &= \mathbb{1}[\text{LLM}(p,x_i)=y_i],\\
s_p(B) &= \tfrac{1}{b}\sum_{i\in B}\operatorname{clip}(u_i(p),[0,C_{\text{clip}}]).
\end{aligned}
\end{equation}
The batch sensitivity is $\Delta_B=C_{\text{clip}}/b$. We release $\tilde{s}_p=s_p+\mathcal{N}(0,(z\Delta_B)^2)$ and account for uniform sampling without replacement at rate $q=b/n$ using R\'enyi DP (RDP), composed over at most $TK$ scoring operations and converted to $(\eps,\delta)$-DP \cite{wang2019subsampled}. The population is kept at exactly $K$ candidates, and all $K$ candidates---including any carried prompts---receive fresh privatized scores in each executed round, giving at most $TK$ releases; early termination can only reduce this count. In the reported optimization runs, $n{=}200$, $b{=}10$, $z{=}10$, $K{=}6$, and $T{=}3$. AutoDP gives $\eps\leq0.71$ at $\delta=10^{-5}$, so all results conservatively satisfy the stated $\eps\leq1$ guarantee. Full accounting details are in Appendix~\ref{sec:appendix:derivations}.

\noindent\textbf{Selection, Composition, End-to-End DP.}
Once the scores are privatized, both deterministic top-$M$ (\textbf{Option B}) and our default Gumbel-smoothed top-$M$ (\textbf{Option A}) are randomized post-processing \cite{dwork2014algorithmic}; neither accesses $\calD$ or incurs additional privacy loss. The Gumbel perturbation uses a fixed, data-independent smoothing scale $\beta$ only to smooth noise-induced rank inversions, not as a second privacy mechanism. Composing at most $TK$ sampled-Gaussian score releases with the RDP accountant therefore establishes end-to-end $(\eps,\delta)$-DP for Algorithm~\ref{alg:dpes} (\refstepcounter{theorem}Theorem~\thetheorem\label{thm:main}; Appendix~\ref{sec:appendix:proofs}).

\begin{figure*}[!t]
\centering
\begin{subfigure}[b]{0.48\textwidth}
    \centering
    \includegraphics[width=\textwidth,height=0.55\textwidth]{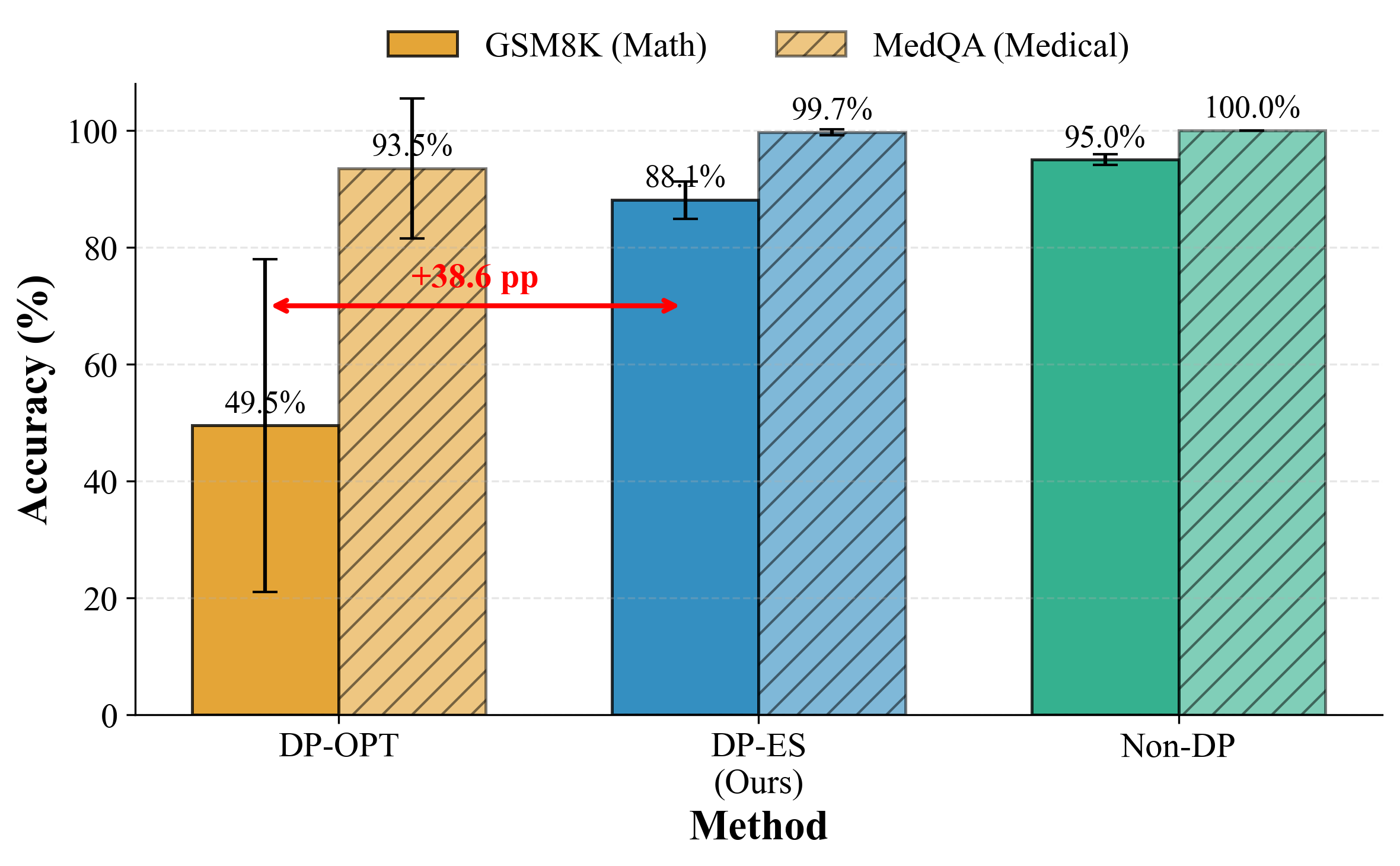}
\end{subfigure}
\hfill
\begin{subfigure}[b]{0.48\textwidth}
    \centering
    \includegraphics[width=\textwidth,height=0.55\textwidth]{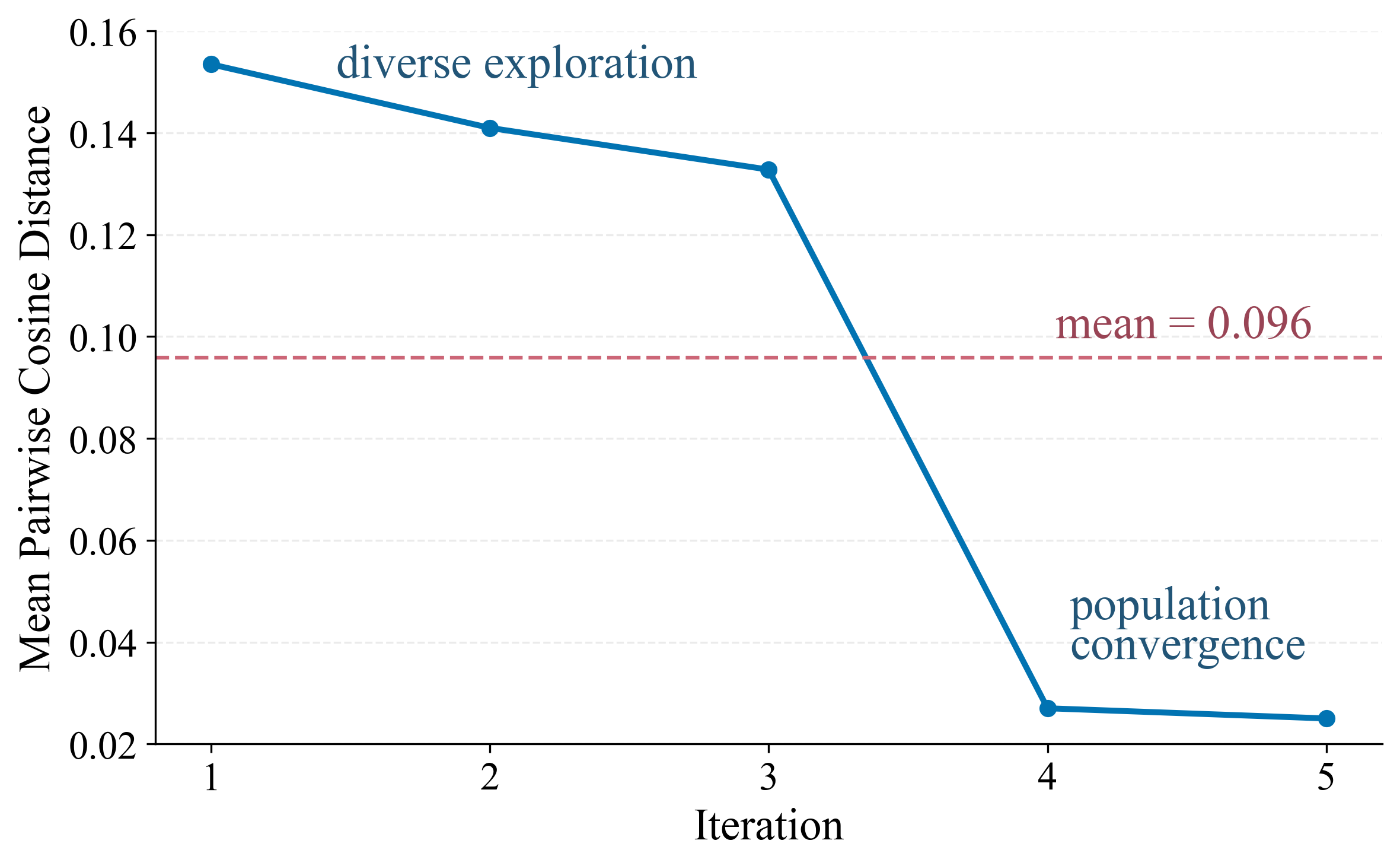}
\end{subfigure}
\caption{\textbf{Performance and search dynamics.} \textbf{Left:} Accuracy on GSM8K and MedQA under the conservative $(\varepsilon{\leq}1,\delta{=}10^{-5})$ guarantee. \textbf{Right:} Mean pairwise prompt-embedding distance, showing early population diversity followed by convergence.}
\label{fig:performance_overview}
\end{figure*}

\section{Experimental Setup}
\label{sec:experiments}

We evaluate DP-ES on four tasks spanning diverse task families: \textbf{GSM8K} \cite{cobbe2021training} (multi-step math reasoning; 200-example optimization set and 1,319-example test set), \textbf{MedQA} \cite{jin2021disease} (medical multiple-choice QA, 200-question subset due to API cost constraints; DP-ES already achieves near-ceiling performance at 99.7\% on this subset, limiting further discriminative power), \textbf{BANKING77} \cite{casanueva2020efficient} (77-class intent classification, 200 examples), and \textbf{Alpaca} \cite{taori2023alpaca} (open-ended instruction-following, 200 examples scored by an LLM judge using DeepSeek-Chat). All experiments use the DeepSeek-V3.2 \cite{liu2024deepseekv3} API. We compare against DP-OPT \cite{hong2024dpopt} and a non-DP baseline---TextGrad \cite{yuksekgonul2024textgrad}. Unless otherwise noted, we fix $K=6$, $M=2$, $T=3$, $b=10$, Gaussian noise multiplier $z=10$, and $\delta=10^{-5}$ across tasks; these settings are not selected using task utility. RDP accounting for uniform sampling without replacement from $n=200$ gives $\varepsilon\leq0.71$, so we report the common conservative guarantee $\varepsilon\leq1$. We also run PromptDPSGD with a local Qwen-2.5-7B model as a soft-prompt tuning baseline; results are summarized in Appendix~\ref{sec:appendix:local}. Full dataset descriptions, hyperparameters, evaluation metrics, and implementation details are provided in Appendix~\ref{sec:appendix:exp}.

\section{Results}
\label{sec:results}

\noindent\textbf{Main Results.}
DP-ES consistently performs well across all four tasks under the common conservative ceiling $(\varepsilon{\leq}1.0,\delta{=}10^{-5})$. Table~\ref{tab:main_results_full} summarizes the results.

\begin{table}[H]
\centering
\scriptsize
\setlength{\tabcolsep}{1.5pt}
\begin{tabular}{lccccc}
\toprule
\textbf{Method} & \textbf{GSM8K} & \textbf{MedQA} & \textbf{BANK77} & \textbf{Alpaca$^\dagger$} & \textbf{Privacy} \\
\midrule
Non-DP & 95.0{\scriptsize$\pm$0.9} & 100.0{\scriptsize$\pm$0.0} & 75.0{\scriptsize$\pm$2.8} & 91.7{\scriptsize$\pm$0.6} & None \\
\midrule
DP-OPT & 49.5{\scriptsize$\pm$28.5} & 93.5{\scriptsize$\pm$12.0} & \textbf{75.3}{\scriptsize$\pm$2.8} & \textbf{87.1}{\scriptsize$\pm$1.4} & (1.0,$10^{-5}$) \\
\textbf{DP-ES} & \textbf{88.1}{\scriptsize$\pm$3.2} & \textbf{99.7}{\scriptsize$\pm$0.5} & 73.5{\scriptsize$\pm$2.2} & 86.8{\scriptsize$\pm$0.8} & ($\leq$0.71,$10^{-5}$) \\
\midrule
\textit{$\Delta$ (ES$-$Opt)} & \textbf{+38.6} & \textbf{+6.2} & $-$1.8 & $-$0.3 & -- \\
\textit{$\Delta$ (ES$-$Non)} & $-$6.9 & $-$0.3 & $-$1.5 & $-$4.9 & -- \\
\bottomrule
\end{tabular}
\caption{\textbf{Main Results} on the task evaluation sets in Section~\ref{sec:experiments} (mean $\pm$ std; 30 runs for GSM8K and three seeds for the other tasks). DP methods satisfy $(\varepsilon{\leq}1.0,\delta{=}10^{-5})$; DP-OPT is reported at its stated $\varepsilon=1$ setting, while DP-ES's recomputed RDP bounds are no larger than 0.71. $^\dagger$Alpaca scores are LLM-judge ratings scaled to percentages.}
\label{tab:main_results_full}
\end{table}
\vspace{-0.8em}

\noindent On GSM8K, DP-ES achieves 88.1\% $\pm$ 3.2\% (+38.6 pp over DP-OPT with $\sim$9$\times$ lower std); on MedQA, 99.7\% $\pm$ 0.5\% (+6.2 pp). On BANKING77 and Alpaca, both methods perform comparably ($\sim$73--75\% and $\sim$87\%, respectively), confirming that DP-ES matches DP-OPT on well-structured tasks while excelling on complex reasoning. Section~\ref{sec:privacy_analysis} separately reports mechanism-level accounting and targeted implementation diagnostics; these do not rely on illustrative prompts as privacy evidence.

\begingroup
\setlength{\textfloatsep}{4pt}
\setlength{\floatsep}{4pt}
\setlength{\intextsep}{4pt}

\begin{figure*}[!t]
\begingroup
\setlength{\textfloatsep}{4pt plus 1pt minus 1pt}
\setlength{\floatsep}{4pt plus 1pt minus 1pt}
\setlength{\intextsep}{4pt plus 1pt minus 1pt}
\captionsetup{font=footnotesize,skip=2pt}
\centering
\vspace{-0.4em}

\begin{tcolorbox}[
    width=0.96\textwidth,
    colback=red!5!white,
    colframe=red!75!black,
    title={\textbf{DP-OPT: Template Drift in the Search Trajectory (GSM8K, real log)}},
    fonttitle=\bfseries,
    sharp corners,
    boxrule=0.4pt,
    before skip=0pt,
    after skip=0pt,
    top=1mm,
    bottom=1mm,
]
\footnotesize
\textbf{Iter 0 (selected):} \textit{``Carefully solve the following math question and provide a clear explanation of your reasoning: \textcolor{green!50!black}{\{question\}}''} \\
\textbf{Iter 1 (selected):} \textit{``Solve the math problem step-by-step, ensuring each step is clearly explained and justified before proceeding to the next.''} \quad \textcolor{red}{\textbf{[\{question\} lost]}} \\
\textbf{Iter 2 (selected, final):} \textit{``Present a structured solution by enumerating each step, explaining the purpose of that step, and confirming its correctness before continuing.''} \quad \textcolor{red}{\textbf{[\{question\} lost]}} \\[0.3em]
\textbf{Trajectory statistics:} 21/33 candidate prompts (64\%) omit \texttt{\{question\}} during search; 2/3 selected prompts in this run omit it. The evaluator appends the question when the placeholder is absent, so this is a structural-drift diagnostic rather than a direct zero-accuracy event.
\end{tcolorbox}

\vspace{0.3em}

\begin{tcolorbox}[
    width=0.96\textwidth,
    colback=orange!5!white,
    colframe=orange!75!black,
    title={\textbf{DP-OPT: Noise-Driven Suboptimal Selection}},
    fonttitle=\bfseries,
    sharp corners,
    boxrule=0.4pt,
    before skip=0pt,
    after skip=0pt,
    top=1mm,
    bottom=1mm,
]
\footnotesize
Noise can reverse the ordering of candidates with similar utility. With no backtracking, an early token-level decision can propagate through subsequent iterations. DP-ES does not eliminate noisy rankings; its population instead retains complete alternatives that can recover in later rounds.
\end{tcolorbox}

\vspace{0.3em}

\begin{tcolorbox}[
    width=0.96\textwidth,
    colback=green!5!white,
    colframe=green!75!black,
    title={\textbf{DP-ES: Complete-Prompt Population Search}},
    fonttitle=\bfseries,
    sharp corners,
    boxrule=0.4pt,
    before skip=0pt,
    after skip=0pt,
    top=1mm,
    bottom=1mm,
]
\footnotesize
\textbf{Final prompt on GSM8K (real log, $\varepsilon{\leq}1.0$):} \textit{``You are a math tutor. Solve the question carefully and show reasoning: \textcolor{green!50!black}{\{question\}}\, (Provide detailed reasoning and justification.) (Be concise and accurate.)''} \quad \textcolor{green!50!black}{\textbf{[\{question\} preserved]}} \\[0.3em]
\textbf{Mechanism:} Prompt-level mutations operate on complete strings; population diversity (6 prompts/round) supplies alternatives after noise-induced rank inversions. \textbf{Result:} $88.1{\pm}3.2\%$ versus DP-OPT's $49.5{\pm}28.5\%$ across 30 runs.
\end{tcolorbox}

\vspace{0.8em}
\captionsetup{aboveskip=2pt,belowskip=0pt}
\caption{\textbf{DP-OPT trajectory drift vs.\ DP-ES population search on GSM8K.} Top: selected prompts in an actual high-accuracy DP-OPT run (seed~42) omit the \texttt{\{question\}} placeholder as search progresses; the harness fallback appends the question, so omission is not treated as an accuracy-failure label. Middle: why irreversible noisy choices can redirect token-level construction. Bottom: a logged DP-ES prompt retains the placeholder while complete alternatives persist across rounds.}
\label{fig:dpopt_failures_summary}
\endgroup
\vspace{-0.5em}
\end{figure*}

\endgroup

\noindent\textbf{DP-OPT Instability and Trajectory Analysis.}
Across 30 GSM8K runs, DP-OPT obtains $49.5{\pm}28.5\%$, compared with DP-ES's $88.1{\pm}3.2\%$. Figure~\ref{fig:dpopt_failures_summary} reproduces a diagnostic DP-OPT trajectory from a \emph{high-accuracy} seed (91\%): 21/33 candidate prompts and 2/3 selected prompts omit \texttt{\{question\}}. The evaluator explicitly appends the question text when this placeholder is absent, so omission is not intrinsically a zero-accuracy event; we therefore treat it as evidence of \emph{template drift}, not as a failure label. The trajectory nonetheless exposes a structural weakness of unconstrained token-level construction, while DP-ES's complete-prompt population retains multiple alternatives when noisy scores perturb a ranking (Appendix~\ref{sec:appendix:dpopt_failures}).

\noindent\textbf{Efficiency.}
DP-ES is \textbf{2.5$\times$ faster} (17.9 vs.\ 45.7 hours on GSM8K) and logs \textbf{3.3$\times$ fewer private-data evaluation/mutation call groups} (10 vs.\ 33 per representative run): DP-OPT's token construction and evaluation both query private aggregates, whereas DP-ES mutations do not touch $\calD$. DP-ES uses larger contexts ($\sim$2.8$\times$ more tokens), so the advantage is lower wall-clock time and fewer private-data round trips rather than uniformly lower provider cost. Full call-purpose and token breakdowns are in Appendix~\ref{sec:appendix:efficiency}.

\noindent\textbf{Local Validation (Qwen2.5-7B).}
To ensure the main trend is not API-specific, we compare all methods locally with Qwen2.5-7B at the common $\varepsilon\leq1$ setting. DP-ES reaches $53.8\pm6.1$\% on GSM8K versus DP-OPT's $37.4\pm11.6$\%; on MedQA the methods are comparable ($81.2\pm1.3$\% vs.\ $83.0\pm2.1$\%). Full results are in Appendix~\ref{sec:appendix:local}.

\noindent\textbf{Why Evolution Strategies Excel Under DP.}
Private Evolution \cite{lin2024dpsda_images,xie2024dpsda_text,lin2025simpe} restricts DP spending to low-sensitivity evaluation while keeping exploration privacy-free. DP-ES inherits this benefit: mutation can be arbitrarily rich without paying privacy budget, enabling recovery from noise-induced regressions. Figure~\ref{fig:performance_overview} (right) shows population diversity (cosine distance $\approx0.15$) in early iterations before convergence, giving DP-ES multiple hypotheses when Gaussian noise perturbs scores.

\noindent\textbf{Selection Ablation: Deterministic vs.\ Gumbel-Smoothed Top-$M$.}\label{sec:em_ablation}
Both selectors operate only on privatized scores and therefore have identical privacy guarantees. Table~\ref{tab:em_ablation} compares them at the default smoothing setting on GSM8K (200 examples).

\begin{table}[t]
\centering
\scriptsize
\setlength{\tabcolsep}{2.5pt}
\begin{tabular}{lcc}
\toprule
\textbf{Selector} & \textbf{Accuracy (\%)} & \textbf{Privacy} \\
\midrule
Gumbel-smoothed & $88.8\pm3.4$ & post-processing \\
Deterministic & $88.6\pm2.9$ & post-processing \\
\bottomrule
\end{tabular}
\caption{\textbf{Selection ablation} (five seeds). The two post-processing rules have comparable accuracy; neither changes the RDP bound established by the Gaussian scores.}
\label{tab:em_ablation}
\end{table}

\noindent The difference in mean accuracy is 0.2 pp and there is no consistent variance advantage at this setting. We retain Gumbel smoothing as the default randomized tie/rank smoother; deterministic top-$M$ is an equally private alternative. Population-size ablations are in Appendix~\ref{sec:appendix:results}.

\section{Privacy Guarantees}
\label{sec:privacy_analysis}

DP-ES inherits formal $(\varepsilon,\delta)$-DP from a \textit{Decoupled DP Evolution} view: mutations operate on public or previously privatized prompts, and only sampled-Gaussian evaluation accesses $\calD$. RDP accounting composes at most $TK$ score releases; both deterministic and Gumbel-smoothed selection are post-processing. The worst reported bound is $(\varepsilon{=}0.71,\delta{=}10^{-5})$, which we conservatively state as $\varepsilon\leq1$. Full derivation and reproducible accountant settings are in Appendix~\ref{sec:appendix:privacy_details}.

\noindent\textbf{Implementation Diagnostics.}
We record 1,000 injected-noise samples and verify that their standardized distribution is consistent with the configured Gaussian mechanism via a Kolmogorov--Smirnov test. We also compare final-prompt behavior on a small set of neighboring datasets as a debugging diagnostic. Because finite output samples cannot certify DP or reliably estimate privacy loss in this high-dimensional discrete space, we do not interpret this diagnostic as an empirical $\varepsilon$; the guarantee comes from the mechanism-level RDP accountant.

\noindent\textbf{Synthetic Privacy Stress Test.}
To illustrate direct memorization risk, we construct 200 synthetic customer profiles containing 1,200 sensitive strings (six per profile: name, phone number, email, full address, street-level address fragment, and account ID). Figure~\ref{fig:privacy_validation} reports exact and near-verbatim matching in final prompts: no listed string is found for DP-ES, while an intentionally adversarial non-DP control---which is explicitly given raw records and instructed to embed a directory---contains 40 strings (3.33\%). This is a targeted sanity check, not a semantic-leakage attack or a substitute for the formal guarantee.


\begin{figure*}[!t]
\centering
\small

\begin{tcolorbox}[
    width=\textwidth,
    colback=blue!5!white,
    colframe=blue!75!black,
    title={\textbf{Experimental Setup: Synthetic Personal Dataset}},
    fonttitle=\bfseries\footnotesize,
    sharp corners,
    boxrule=0.4pt,
    left=3pt,right=3pt,top=3pt,bottom=3pt
]
\footnotesize
\textbf{Dataset:} 200 synthetic customer profiles, each yielding six sensitive strings: full name, phone number, email address, full address, street-level address fragment, and account ID. \textbf{1,200 unique sensitive strings total.}

\vspace{0.1em}
\textbf{Evaluation:} Rewards answers that recover profile details. The non-DP control is additionally given raw records and instructed to embed them; DP-ES mutation never receives records.

\vspace{0.1em}
\textbf{Task:} ``Generate a customer contact information query prompt''
\end{tcolorbox}

\vspace{0.3em}

\begin{minipage}[t]{0.48\textwidth}
\begin{tcolorbox}[
    colback=red!5!white,
    colframe=red!75!black,
    title={\textbf{Non-DP: Privacy Leakage}},
    fonttitle=\bfseries\footnotesize,
    sharp corners,
    boxrule=0.4pt,
    left=3pt,right=3pt,top=3pt,bottom=3pt
]
\footnotesize
\textbf{Final Prompt (excerpt):}
\begin{quote}
\scriptsize\ttfamily
Use this customer directory:\\
\textcolor{red}{\textbf{Jessica Moore}}, (713) 907-8800,\\
\textcolor{red}{\textbf{jessica.moore83@mail.com}};\\
\textcolor{red}{\textbf{Anthony Williams}}, (312) 991-9095,\\
\textcolor{red}{\textbf{anthony.williams60@mail.com}}; \ldots
\end{quote}

\vspace{0.1em}
\textbf{Privacy Analysis:}
\begin{itemize}[leftmargin=*, itemsep=0pt, parsep=0pt, topsep=1pt]
    \item \textcolor{red}{\textbf{Leaked: 40 strings (3.33\% of 1,200)}}
    \item Embeds names, phones, emails, addresses, and account IDs
\end{itemize}
\end{tcolorbox}
\end{minipage}
\hfill
\begin{minipage}[t]{0.48\textwidth}
\begin{tcolorbox}[
    colback=green!5!white,
    colframe=green!75!black,
    title={\textbf{DP-ES Configuration}},
    fonttitle=\bfseries\footnotesize,
    sharp corners,
    boxrule=0.4pt,
    left=3pt,right=3pt,top=3pt,bottom=3pt
]
\footnotesize
\textbf{Final Prompt:}
\begin{quote}
\scriptsize\ttfamily
You are a customer service assistant.\\
Retrieve relevant information based on\\
user queries. If the query involves\\
account information, verify before responding.
\end{quote}

\vspace{0.1em}
\textbf{Privacy Analysis:}
\begin{itemize}[leftmargin=*, itemsep=0pt, parsep=0pt, topsep=1pt]
    \item \textcolor{green!50!black}{\textbf{Leaked: 0 strings (0.0\% of 1,200)}}
    \item Generic instructions; no listed PII string detected
\end{itemize}
\end{tcolorbox}
\end{minipage}

\vspace{0.2em}

\caption{\textbf{Exact-match memorization stress test} on 200 synthetic profiles (1,200 sensitive strings). The intentionally adversarial \textbf{Non-DP} control is shown raw records and asked to embed a directory; its final prompt contains 40 strings (3.33\%). DP-ES mutation never sees records, and its final prompt contains none of the listed strings. Excerpts and counts are reproduced from the saved run. This targeted diagnostic does not test semantic leakage or replace the formal guarantee.}
\label{fig:privacy_validation}
\end{figure*}

\section{Conclusion}
\label{sec:conclusion}

We presented a \emph{diagnostic and methodological} contribution for differentially private prompt optimization. On the diagnostic side, a 30-run analysis shows substantial DP-OPT instability on GSM8K, while a logged trajectory exposes prompt-template drift and noise-sensitive irreversible decisions in token-level construction. On the methodological side, we introduced \textbf{DP-ES}, a structurally cleaner evolution-based alternative that maintains a population of full prompts, mutates them without exposing private records, and restricts privacy expenditure to sampled-Gaussian scoring; selection over privatized scores is post-processing.

Across four tasks---GSM8K (88.1\%), MedQA (99.7\%), BANKING77 (73.5\%), and Alpaca (86.8\%, LLM-judge)---DP-ES matches or exceeds DP-OPT under a conservative $\varepsilon\leq1$ guarantee, with $\sim$9$\times$ lower variance on GSM8K, 2.5$\times$ faster wall-clock runtime, and 3.3$\times$ fewer logged private-data call groups. Deterministic and Gumbel-smoothed selection have comparable utility and identical privacy guarantees because both post-process privatized scores. Noise-distribution checks and a 200-profile exact-match memorization stress test provide complementary diagnostics alongside the formal RDP accounting.

\noindent\textbf{Scope.} The strongest evidence we provide is for \emph{optimization robustness under DP noise on benchmarks where prompt structure matters}, with the largest gains on reasoning-heavy tasks where token-level construction is most brittle. End-to-end validation on genuinely sensitive, non-saturated deployment data---which would require benchmarks combining record-level sensitivity with non-trivial semantic reasoning---remains future work and is the most direct extension of this contribution.

\section*{Limitations}

We are explicit about the boundaries of what this work establishes.

\noindent \textbf{Empirical Scope: No Sensitive-Data Deployment Validation.}
Our experiments establish that DP-ES is a structurally more robust DP prompt optimizer than token-level alternatives on benchmarks where prompt structure matters. They do \textit{not} validate end-to-end deployment on genuinely sensitive, non-saturated datasets (e.g., real clinical records, financial logs). Such validation would require benchmarks that combine record-level sensitivity with non-trivial semantic reasoning, which the current public benchmark landscape does not provide. The formal $(\varepsilon,\delta)$-DP guarantee, by construction, bounds record-level leakage independent of the benchmark, but a direct sensitive-data deployment study remains future work.

\noindent \textbf{MedQA Ceiling Effect.}
DP-ES reaches 99.7\% on the 200-question MedQA subset, with the Non-DP baseline saturating at 100.0$\pm$0.0. This near-ceiling performance limits the discriminative power of MedQA for studying privacy-utility trade-offs at $\varepsilon{=}1.0$; the 6.2~pp improvement over DP-OPT on this saturated regime is informative but should not be over-interpreted as evidence for a non-trivial privacy-utility frontier on medical QA more broadly. A non-saturated medical benchmark (e.g., a harder open-ended clinical reasoning task) would provide stronger evidence.

\noindent \textbf{Privacy Attack Coverage.}
The 200-profile synthetic memorization stress test (Section~\ref{sec:privacy_analysis}) measures direct or near-verbatim sensitive-string leakage against an intentionally adversarial non-DP control. It does \textit{not} test paraphrased memorization, rare-attribute disclosure, record reconstruction, or membership inference. Moreover, the mutation LLM can reproduce information already present in its parametric memory; DP-ES bounds the incremental influence of the private optimization dataset, not memorization inherited from pretraining. The formal $(\varepsilon,\delta)$-DP guarantee is attack-agnostic, but a direct empirical attack suite comparing DP-ES against DP-OPT under matched bounds would provide useful complementary evidence and remains future work.

\noindent \textbf{Efficiency Reporting Scope.}
The cost decomposition by call purpose (Appendix~\ref{sec:appendix:efficiency}, Table~\ref{tab:efficiency_breakdown}) is from a representative single-seed run on a 200-sample validation set. Aggregate wall-clock numbers across the full 1{,}319-sample test set are reported separately (Table~\ref{tab:efficiency_full}); token-level and call-purpose splits at full aggregate scale are not yet released. We report the most granular reproducible numbers we have and treat aggregate token-level reporting as future work.

\noindent \textbf{Task Diversity.}
We evaluate on four tasks (GSM8K, MedQA, BANKING77, Alpaca) covering math reasoning, medical QA, intent classification, and instruction-following. Additional benchmarks (e.g., summarization, code generation, open-ended clinical reasoning) would further characterize cross-domain generalization.

\noindent \textbf{Task-Dependent Advantage.}
DP-ES's strongest gains over DP-OPT are on complex reasoning tasks (GSM8K: +38.6 pp, MedQA: +6.2 pp). On well-structured classification (BANKING77) and instruction-following (Alpaca), both methods perform comparably, suggesting that DP-ES's population-based exploration is most beneficial when the prompt search space is large and structured prompts are critical.

\noindent \textbf{Privacy Budget Sweeps.}
Our verified experiments establish the common conservative bound $\varepsilon\leq1$ rather than a densely sampled privacy--utility frontier. Re-running all methods with noise recalibrated to several exact RDP targets would better characterize the trade-off.

\noindent \textbf{Theoretical Analysis of Task Complexity.}
We empirically observe that task complexity is a key determinant of the success of DP optimization. A formal theoretical framework linking prompt search space characteristics to privacy budget requirements is still needed.

\noindent \textbf{Future Work.}
Beyond the scope items above, promising directions include adaptive population sizing, multi-objective optimization (utility, privacy, stability), and federated DP-ES for distributed sensitive data settings.

\section*{Ethics Statement}

This work develops differentially private (DP) prompt-optimization mechanisms motivated by settings where the underlying optimization data may be sensitive. Our empirical evaluation is conducted on public benchmarks; the considerations below concern both this study and any downstream use of the released method.

\noindent \textbf{Privacy Scope:} DP-ES provides formal ($\varepsilon, \delta$)-DP guarantees for prompt optimization under correct implementation and accounting. The reported implementation diagnostics and exact-match stress test do not certify DP or cover downstream deployment risks beyond the mechanism.

\noindent \textbf{Data Use:} We use public benchmarks and do not collect new personal data. Applications to real sensitive data should follow institutional review, access control, and data governance requirements.

\noindent \textbf{Dual-Use Concerns:} As with any optimization technique, DP-ES could be misused to refine harmful applications. Responsible deployment should include access control, monitoring, and safety filtering.

\noindent \textbf{Transparency and Reproducibility:} We provide code and privacy accounting details to enable verification of claims and independent audits.

\section*{Acknowledgments}

OpenAI Codex was used to assist with language polishing, manuscript organization, code debugging, and consistency checks. The authors independently verified all technical claims, calculations, citations, and final wording and take full responsibility for the paper.

\bibliography{references}

\clearpage
\appendix
\section{Experimental Setup}
\label{sec:appendix:exp}

\noindent This appendix provides comprehensive experimental setup details, including tasks, baselines, hyperparameters, and implementation specifics.
\setlist[itemize]{itemsep=1pt, topsep=1pt, parsep=0pt, partopsep=0pt}
\setlist[enumerate]{itemsep=1pt, topsep=1pt, parsep=0pt, partopsep=0pt}

\subsection{Datasets}

\noindent \textbf{GSM8K} \cite{cobbe2021training}: Grade school math word problems requiring multi-step arithmetic reasoning. We use the full official test set of 1,319 examples from the original benchmark for final evaluation. During prompt optimization, we use a separate 200-example validation subset to compute fitness scores, ensuring that no test data leakage occurs. Problems involve operations like addition, multiplication, and percentages. Success requires prompts that guide the step-by-step generation of solutions.

\noindent \textbf{MedQA} \cite{jin2021disease}: Medical question answering from the USMLE exam. We use 200 questions covering disease diagnosis, treatment, and pathophysiology. Evaluation uses choice-only accuracy on this 200-question subset; we do not fine-tune model weights (prompt-only optimization), avoiding training-set leakage. This task probes performance on a specialized domain and high-accuracy regimes.

\noindent \textbf{BANKING77} \cite{casanueva2020efficient}: Fine-grained intent classification with 77 banking-related intent categories (e.g., \texttt{card\_arrival}, \texttt{exchange\_rate}, \texttt{lost\_or\_stolen\_card}). We use 200 randomly sampled examples for evaluation. This task tests DP-ES on a classification problem with a large label space, where prompt optimization must convey category structure without memorizing individual examples.

\noindent \textbf{Alpaca} \cite{taori2023alpaca}: Open-ended instruction-following evaluated by an LLM judge. We use 200 instructions sampled from the Alpaca evaluation set. Quality is scored on a $[0,1]$ rubric by DeepSeek-Chat (temperature 0.0, max 64 tokens), assessing instruction adherence, factual accuracy, and response quality. This task represents a fundamentally different evaluation paradigm (generative quality via LLM-judge rather than exact-match accuracy), testing whether DP-ES generalizes beyond closed-form metrics.

\subsection{Baselines}

\textbf{DP-OPT (black-box adaptation)} \cite{hong2024dpopt}: DP-OPT's canonical method uses histogram-based token generation with the exponential mechanism. Because the original implementation targets a white-box local model, our API comparison uses a black-box adaptation that preserves private candidate evaluation and exponential-mechanism selection but is not an exact reproduction of the original Vicuna-based implementation. We provide the original implementation link for reference\footnote{\url{https://github.com/VITA-Group/DP-OPT}} and interpret the resulting numbers as an adapted baseline.

\noindent \textbf{Non-DP}: Standard prompt optimization without privacy constraints, implemented using the TextGrad\footnote{\url{https://github.com/zou-group/textgrad}} \cite{yuksekgonul2024textgrad}
 method. This serves as a non-private reference rather than a formal upper bound, since optimization procedures can differ in search behavior and local optima.

\noindent \textbf{PromptDPSGD (local)} \cite{Duan2023FlocksOS}: DP-SGD on soft prompt embeddings. We run PromptDPSGD with local Qwen-2.5-7B, prompt length 20, 10 epochs, learning rate 0.005, max grad norm 1.0, batch size 2 or 4. Results are summarized in Appendix~\ref{sec:appendix:local}.

\subsection{Hyperparameters}

\textbf{DP-ES configuration:}
\begin{itemize}
    \item Population size: $K=6$ prompts per generation; $M=2$ parents selected per round
    \item Iterations: $T=3$; sampled scoring batch $b=10$
    \item Gaussian mechanism: score sensitivity $1/b$, noise standard deviation $1.0$ (noise multiplier $z=10$)
    \item Selection: Gumbel-smoothed top-$M$ over privatized scores (default); deterministic top-$M$ is an alternative
    \item Accounting: uniform subsampling without replacement and RDP composition over at most $TK=18$ score releases, converted at $\delta=10^{-5}$
    \item Mutation mode: BALANCED (mix of exploration and exploitation)
    \item Randomness: run counts are reported with each experiment; the three-seed validation studies use seeds [42, 123, 456]
\end{itemize}

\noindent \textbf{DP-OPT configuration:} We use 6 candidate prompts per iteration, 3 iterations, and its reported $(\varepsilon{=}1,\delta{=}10^{-5})$ setting. DP-ES is compared under the same conservative ceiling; its recomputed bound is tighter ($\varepsilon\leq0.71$).

\noindent \textbf{Initial prompt:} For both methods, we use a generic starting prompt: \textit{``Answer the following question carefully.''}





\subsection{Implementation}

\textbf{LLM API:} We use DeepSeek-V3.2 (deepSeek-chat) via the OpenAI-compatible API\footnote{\url{https://api.deepseek.com/v1}}.
    
\noindent \textbf{Privacy accounting:} We use the \texttt{autodp} library \cite{autodp} with the R\'enyi DP accountant for tight composition bounds.

\noindent \textbf{Reproducibility:} We report run counts with each result (30 runs for the GSM8K stability analysis, five seeds for the selector ablation, and three seeds for the validation studies). Code, prompts, accountant inputs, and evaluation scripts accompany the final paper.


\subsection{Mutation Strategies and Templates}
\textbf{Mutation strategies:} We implement three modes:
\begin{itemize}
    \item \texttt{EXPLORE}: Generate semantically diverse variations
    \item \texttt{EXPLOIT}: Make small refinements to successful prompts
    \item \texttt{BALANCED}: Mix exploration and exploitation
\end{itemize}

\noindent The mutation prompt template is:
\begin{quote}
\textit{``You are a prompt optimization assistant. Generate [K] creative variations of the following prompt that might improve performance: [parent prompt]. Make [large/small] semantic changes.''}
\end{quote}

\section{Privacy Guarantees and Verification}
\label{sec:appendix:privacy_details}

\subsection{Theoretical Framework: Decoupled DP Evolution}
We restate the privacy argument in full. Let $\calD$ be the private dataset and $\calP_1$ an initial population generated without access to $\calD$. For $t=1,\dots,T$ define

\begin{align*}
  Z_t &= \mathcal{Q}_t(\calD, \calP_t), &
  \mathcal{S}_t &= \mathrm{Sel}_t(Z_t), \\
  \calP_{t+1} &= G_t(\mathcal{S}_t, U_t). &&
\end{align*}

\noindent where $\mathcal{Q}_t$ releases $K$ sampled-Gaussian scores in each executed round, $\mathrm{Sel}_t$ retains $M$ parents from those privatized scores, and $G_t$ mutates selected prompts using only public randomness $U_t$. Both $\mathrm{Sel}_t$ and $G_t$ are post-processing. Thus the total privacy cost is that of at most $TK$ adaptively composed sampled-Gaussian score releases.

\begin{proof}[Proof sketch]
\textbf{Step 1 (Privacy source).} At each round $t$, only $\mathcal{Q}_t$ touches $\calD$. Each candidate score uses a uniformly sampled batch of size $b$, clips per-record utility to $[0,1]$, averages it, and adds Gaussian noise with standard deviation $z/b$.

\noindent \textbf{Step 2 (Post-processing).} Selection sees only privatized scores; mutation sees only public or previously privatized prompts. Neither operation adds privacy loss.

\noindent \textbf{Step 3 (RDP composition).} We apply the without-replacement subsampling bound of \citet{wang2019subsampled} to each score release, add RDP across at most $TK$ adaptive releases, and convert to $(\varepsilon,\delta)$-DP. Post-processing then gives the same guarantee for the released final prompt.

\textbf{Intuition.} Private data influence the search only through noisy scalar scores. Every later operation---ranking, randomized smoothing, LLM mutation, and final choice---uses already privatized outputs.
\end{proof}

\subsection{Supplementary Derivations}
\label{sec:appendix:derivations}

\noindent\textbf{Sampled-Gaussian accounting.}
For a batch mean with clipped per-record utility in $[0,1]$, replace-one sensitivity is $\Delta_B=1/b$. Adding noise $\mathcal{N}(0,(z/b)^2)$ is a Gaussian mechanism with noise multiplier $z$. We apply uniform subsampling without replacement at rate $q=b/n$, compose at most $TK$ mechanisms in RDP, and convert at $\delta=10^{-5}$ using AutoDP \cite{autodp}. Reported runs use $b=10$ and $z=10$. With $K=6$ and $T=3$, the resulting worst-case bound is given in Table~\ref{tab:rdp_bounds}.

\begin{table*}[t]
\centering
\small
\begin{tabular}{lccc}
\toprule
\textbf{Setting} & $n$ & $q$ & $\varepsilon$ at $\delta=10^{-5}$ \\
\midrule
All reported sampled-score runs & 200 & 0.05 & $\leq0.71$ \\
\bottomrule
\end{tabular}
\caption{Recomputed worst-case privacy bound for at most 18 sampled-Gaussian score releases in the reported DP-ES configuration. We report the common conservative ceiling $\varepsilon\leq1$.}
\label{tab:rdp_bounds}
\end{table*}

\noindent\textbf{Gumbel-smoothed selection.}
For smoothing only, we add independent Gumbel noise to each privatized score before ranking, with $\beta$ fixed independently of $\calD$:
\begin{equation}
\hat{s}_p = \tilde{s}_p + \operatorname{Gumbel}(0,\beta).
\end{equation}
Because this operation is a function only of already privatized scores and fresh randomness, it is post-processing for every $\beta\geq0$. Deterministic top-$M$ is the $\beta=0$ special case. Table~\ref{tab:em_ablation} compares their utility.

\subsection{Comparison with DP-OPT Privacy}

\noindent DP-OPT is evaluated at its reported $(\varepsilon{=}1.0,\delta{=}10^{-5})$ setting; DP-ES satisfies the same ceiling with a tighter recomputed bound. DP-ES achieves:

\begin{itemize}
    \item \textbf{Better utility on complex tasks:} +38.6 pp on GSM8K and +6.2 pp on MedQA under the common $\varepsilon\leq1$ ceiling
    \item \textbf{Competitive on simpler tasks:} Within 2 pp of DP-OPT on BANKING77 and Alpaca
    \item \textbf{Decoupled exploration:} Only sampled-Gaussian scores consume privacy; mutation and selection are post-processing
    \item \textbf{Reproducible accounting:} The released accountant inputs yield $\varepsilon\leq0.71$ for all reported DP-ES tasks
\end{itemize}

This comparison shows that changing the search structure can improve utility while satisfying an equal or stronger privacy bound.

\subsection{Threat Model and Assumptions}

\noindent Our privacy guarantees assume:

\begin{enumerate}
    \item \textbf{Trusted curator:} The party running DP-ES has access to the private dataset $\calD$ but releases only the final prompt with DP guarantees.

    \item \textbf{Record-level privacy:} The privacy unit is one data record $(x_i, y_i)$. Group privacy (protecting multiple records per individual) requires scaling $\varepsilon$ by group size.

    \item \textbf{Trusted computation boundary:} The curator and any model endpoint used for private-data evaluation are trusted processors. The released prompt is protected; DP does not hide API inputs from the provider receiving them.

    \item \textbf{Mutation isolation:} The mutation LLM receives only a parent prompt, never records or record-derived text feedback. PII memorized independently in the pretrained model's parameters is outside this record-level threat model and should be handled by provider safeguards.

    \item \textbf{Post-processing safety:} The released prompt may be used by downstream applications without additional loss about $\calD$, although new downstream data require their own privacy analysis.
\end{enumerate}

These assumptions are standard in the DP literature and match those of DP-OPT and other DP-ML methods.

\subsection{Implementation Diagnostics}

We log the exact noise draws used by the scorer. For both GSM8K and MedQA, 1,000 standardized samples are consistent with the configured Gaussian distribution under a Kolmogorov--Smirnov check. We also rerun the optimizer on five neighboring datasets as a debugging exercise. The resulting nearest-neighbor output-distance statistic is degenerate on near-ceiling MedQA and is not a calibrated likelihood-ratio estimator; accordingly, we do not report it as an observed privacy parameter. These checks can catch missing noise or gross data-flow errors, but only the accountant in Table~\ref{tab:rdp_bounds} establishes the formal guarantee.

\section{Additional Results and Analyses}
\label{sec:appendix:results}

\noindent Main results appear in Section~\ref{sec:results}. This appendix provides efficiency comparisons, DP-OPT stability and trajectory analysis, local validation (including PromptDPSGD), ablations, accounting details, and implementation diagnostics.

\subsection{Efficiency Comparison}
\label{sec:appendix:efficiency}

\noindent Table~\ref{tab:efficiency_full} presents detailed runtime comparisons between DP-ES and DP-OPT across both API and local deployments.

\begin{table*}[t]
\centering
\small
\begin{tabular}{lccc}
\toprule
\textbf{Task} & \textbf{DP-OPT} & \textbf{DP-ES (Ours)} & \textbf{Speedup} \\
\midrule
\multicolumn{4}{l}{\textit{DeepSeek API ($\varepsilon\leq1.0$, 1,319 test examples, 3 iterations):}} \\
GSM8K & 45.7 hours & \textbf{17.9 hours} & \textbf{2.5$\times$} \\
MedQA & 18.4 hours & \textbf{7.0 hours} & \textbf{2.6$\times$} \\
\midrule
\multicolumn{4}{l}{\textit{Local Qwen2.5-7B ($\varepsilon\leq1.0$, 200 examples):}} \\
GSM8K & 14.60 hours & \textbf{7.37 hours} & \textbf{1.98$\times$} \\
MedQA & 3.24 hours & \textbf{1.87 hours} & \textbf{1.73$\times$} \\
\bottomrule
\end{tabular}
\caption{\textbf{Efficiency Comparison (wall-clock).} DP-ES is consistently faster than DP-OPT across both API and local deployments due to efficient population-based evaluation.}
\label{tab:efficiency_full}
\end{table*}

\noindent\textbf{Cost Breakdown by Call Purpose.}
Wall-clock runtime alone does not isolate \textit{where} the efficiency gain comes from. Table~\ref{tab:efficiency_breakdown} decomposes a representative GSM8K run (seed~42, 200-example optimization set, $\varepsilon\leq1.0$) by call purpose, separating mutation from private-data evaluation. Two observations:
\begin{itemize}[leftmargin=*,itemsep=-0.2em,topsep=2pt]
\item \textbf{DP-OPT's mutation calls also consume the DP budget.} DP-OPT's histogram-based token construction queries private counts at every token, so its 3 mutation calls (90 token-level votes in total) are not free---they enter the privacy accountant alongside the 30 evaluation calls.
\item \textbf{DP-ES mutation is post-processing.} DP-ES generates whole-prompt mutations via LLM calls that never touch the private dataset; only its evaluation call groups access private records. Although DP-ES uses larger contexts, the number of private-data call groups is 3$\times$ smaller.
\end{itemize}
The resulting efficiency profile is best described as \textit{trade tokens for robustness}: DP-ES uses privacy-free mutation and fewer private-data call groups. Wall-clock runtime drops by 2.5$\times$ because fewer serial round-trips dominate latency, not because token count is lower.

\begin{table*}[t]
\centering
\small
\setlength{\tabcolsep}{4pt}
\begin{tabular}{lrrr}
\toprule
\textbf{Cost component} & \textbf{DP-OPT} & \textbf{DP-ES} & \textbf{Ratio} \\
\midrule
\multicolumn{4}{l}{\textit{Logged private-data call groups:}} \\
\quad Mutation (token-level / privacy-free) & 3 (DP) & 0 (free) & --- \\
\quad Evaluation (Gaussian-noised) & 30 & 10 & 3.0$\times$ fewer \\
\quad \textbf{Total private-data groups} & \textbf{33} & \textbf{10} & \textbf{3.3$\times$ fewer} \\
\midrule
\multicolumn{4}{l}{\textit{Token usage:}} \\
\quad Input tokens & 1{,}309 & 6{,}120 & 4.7$\times$ more \\
\quad Output tokens & 2{,}707 & 5{,}000 & 1.8$\times$ more \\
\quad Total tokens & 4{,}016 & 11{,}120 & 2.8$\times$ more \\
\midrule
\multicolumn{4}{l}{\textit{Wall-clock:}} \\
\quad Runtime (hours) & 16.6 & 6.5 & 2.5$\times$ faster \\
\quad Accuracy (single run) & 91.0\% & 86.5\% & --- \\
\bottomrule
\end{tabular}
\caption{\textbf{Efficiency breakdown by call purpose} (GSM8K, seed 42, $\varepsilon\leq1$, 200-example optimization set). Counts are logger-level call groups, not individual per-example model requests. DP-ES uses more tokens but fewer private-data round trips, yielding 2.5$\times$ lower wall-clock time in this setting. Provider pricing and validation size can change the cost comparison.}
\label{tab:efficiency_breakdown}
\end{table*}

\subsection{DP-OPT Stability and Trajectory Analysis}
\label{sec:appendix:dpopt_failures}

\noindent Across 30 GSM8K runs, DP-OPT achieves $49.5{\pm}28.5\%$, whereas DP-ES achieves $88.1{\pm}3.2\%$. We use these distributional statistics directly rather than imposing a post-hoc accuracy threshold to label runs as failures. Table~\ref{tab:dpopt_stability} summarizes the aggregate statistics and representative trajectory diagnostic. The main-text visualization (Figure~\ref{fig:dpopt_failures_summary}) reproduces an actual DP-OPT trajectory and illustrates two mechanisms consistent with the observed instability: prompt-template drift during token-level construction and irreversible propagation of noisy ranking decisions.

\begin{table*}[t]
\centering
\small
\begin{tabular}{lccc}
\toprule
\textbf{Method / diagnostic} & \textbf{Runs} & \textbf{Accuracy} & \textbf{Observed structure} \\
\midrule
DP-OPT aggregate & 30 & $49.5{\pm}28.5\%$ & high run-to-run variance \\
DP-ES aggregate & 30 & $88.1{\pm}3.2\%$ & lower run-to-run variance \\
DP-OPT seed 42 trajectory & 1 & 91.0\% & 21/33 candidates omit \texttt{\{question\}} \\
\bottomrule
\end{tabular}
\caption{\textbf{DP-OPT stability and trajectory diagnostics.} The seed-42 trajectory is deliberately reported alongside its high final accuracy: the evaluation harness appends the question when \texttt{\{question\}} is absent, so placeholder omission is a structural-drift diagnostic rather than a direct failure criterion.}
\label{tab:dpopt_stability}
\end{table*}

\noindent \textbf{Root Cause Analysis:}

\noindent \textbf{1. Token-Level Myopia:} DP-OPT generates prompts token by token using a histogram-based exponential mechanism. At each position, it votes on the next token across data partitions. This greedy process lacks global coherence checking, allowing it to:
\begin{itemize}
    \item Omit critical placeholders like \texttt{\{question\}} during generation
    \item Create syntactically valid but semantically meaningless prompts
    \item Drift toward generic text unrelated to the task
\end{itemize}

\noindent \textbf{2. Non-Backtracking Search:} Once a token is selected, DP-OPT cannot revise it. If early tokens lead to a poor prompt structure, the algorithm is permanently stuck. This contrasts with DP-ES, which maintains a diverse population and can recover from individual poor candidates.

\noindent \textbf{3. Sample-Based Evaluation Noise:} DP-OPT evaluates candidates on small samples (typically 10-20 examples). A prompt that performs well by chance on this sample may be selected and propagated, even if it generalizes poorly. DP-ES mitigates this through population diversity—multiple independent candidates reduce the risk of overfitting to evaluation noise.
\vspace{0.4em}

\noindent \textbf{Noise-sensitive selection.} DP noise can reverse the ordering of candidates with similar utility. DP-OPT commits to each token-level decision without backtracking, so an early rank inversion can permanently redirect construction. DP-ES does not eliminate rank inversions; instead, its population retains multiple complete hypotheses and can recover in later rounds.
\vspace{0.5em}

\noindent \textbf{Why DP-ES Improves Stability:}

\begin{itemize}
    \item \textbf{Prompt-level generation:} Mutations operate on complete prompts, preserving global structure
    \item \textbf{Population diversity:} Even if one candidate is degraded by noise, others can compensate
    \item \textbf{Privacy-free exploration:} Mutations don't consume privacy budget, allowing extensive search
\end{itemize}

This analysis offers a mechanism-level explanation consistent with the 38.6 pp mean performance gap and the substantially lower variance of DP-ES on GSM8K.

\subsection{Local Validation with Qwen2.5-7B}
\label{sec:appendix:local}

\noindent To check whether the main task-dependent trend transfers beyond the API model, we evaluate four methods with a local Qwen2.5-7B model \cite{qwen2.5} under the common verified $\varepsilon\leq1$ setting. Table~\ref{tab:local_validation} reports the results.

\subsubsection{Experimental Setup}

\noindent\textbf{Model:} Qwen2.5-7B-Instruct (local deployment) \\
\textbf{Methods:} Non-DP, DP-ES, DP-OPT, PromptDPSGD \\
\textbf{Tasks:} GSM8K, MedQA \\
\textbf{Privacy guarantee:} $\varepsilon\leq1$, $\delta=10^{-5}$ \\

\begin{table}[t]
\centering
\small
\setlength{\tabcolsep}{3pt}
\begin{tabular}{@{}lcc@{}}
\toprule
\textbf{Method} & \textbf{GSM8K} & \textbf{MedQA} \\
\midrule
DP-ES & $53.8\pm6.1$ & $81.2\pm1.3$ \\
DP-OPT & $37.4\pm11.6$ & $83.0\pm2.1$ \\
PromptDPSGD & $20.0\pm5.8$ & $34.8\pm19.0$ \\
Non-DP & $26.2\pm10.2$ & $79.5\pm4.2$ \\
\bottomrule
\end{tabular}
\caption{\textbf{Local Qwen2.5-7B validation} at the common $\varepsilon\leq1$ setting (three seeds; mean $\pm$ std). Non-DP is a reference rather than an upper bound because the methods use different search procedures.}
\label{tab:local_validation}
\end{table}

\subsubsection{Key Findings}

\begin{itemize}
    \item \textbf{GSM8K:} DP-ES exceeds DP-OPT by 16.4 pp and halves the standard deviation, reproducing the reasoning-task advantage.
    \item \textbf{MedQA:} DP-ES and DP-OPT are comparable, consistent with the task-dependent pattern in the API results.
    \item \textbf{PromptDPSGD:} The soft-prompt baseline trails both discrete methods on this 7B model.
\end{itemize}

\subsection{Ablation Studies}

\noindent To validate our design choices, we conduct ablations on GSM8K under the common $\varepsilon\leq1$ guarantee.

\noindent \textbf{Population Size Ablation (Table~\ref{tab:pop_ablation}):} We vary population size while keeping $K\times T\leq24$. $K=6,T=3$ gives the best observed diversity--depth balance; smaller populations explore too narrowly, while larger populations use fewer refinement rounds under the fixed search-call cap.

\begin{table}[t]
\centering
\small
\begin{tabular}{cccc}
\toprule
\textbf{Population} & \textbf{Iterations} & \textbf{Accuracy} & \textbf{$\Delta$ vs Baseline} \\
\midrule
3  & 4 & 78.5\% & -8.0 pp \\
\textbf{6} & \textbf{3} & \textbf{86.5\%} & \textbf{--} \\
9  & 2 & 82.0\% & -4.5 pp \\
12 & 2 & 82.0\% & -4.5 pp \\
\bottomrule
\end{tabular}
\caption{\textbf{Population Size Ablation.} $K=6$, $Iterations=3$ provides an optimal diversity-depth tradeoff.}
\label{tab:pop_ablation}
\end{table}

\subsection{Practitioner Hyperparameter Guide}
\label{sec:appendix:hyperparameter_guide}

\noindent Start with $K=6,T=3$ and recompute the RDP bound whenever $K$, $T$, $b$, $n$, or $z$ changes. Increasing $KT$ increases both evaluation cost and composed privacy loss; reducing $K$ narrows exploration, while reducing $T$ limits iterative refinement.

\subsection{Selection Mechanism Ablation (Extended)}
\label{sec:appendix:em_ablation}

\noindent Table~\ref{tab:em_ablation} reports the five-seed comparison at the default setting. Gumbel-smoothed and deterministic top-$M$ are statistically similar (88.8\% vs.\ 88.6\%) and have identical privacy accounting because both post-process privatized scores. We use smoothing by default but do not claim a consistent variance advantage.

\subsection{Population Diversity Analysis}
\label{sec:appendix:diversity}

\noindent To understand DP-ES's exploration-exploitation dynamics, we analyzed population diversity over 5 iterations using sentence-BERT embeddings (all-mpnet-base-v2). We measure diversity as the average pairwise cosine distance between prompt embeddings. Figure~\ref{fig:performance_overview} (right) visualizes this trajectory.

\vspace{-0.4em}
\section{Proofs}
\label{sec:appendix:proofs}

\vspace{-0.3em}
\subsection{Proof of Proposition~\ref{prop:mutation-free}}
By the post-processing property of differential privacy \cite{dwork2014algorithmic}, a function of public inputs or previously privatized outputs adds no privacy loss. Mutate receives only a parent prompt and public randomness, never $\calD$ or record-derived feedback; it is therefore post-processing.

\vspace{-0.3em}
\subsection{Proof of Proposition~\ref{prop:decoupled}}
Because mutation and candidate generation do not access $\calD$, they incur zero additional privacy loss. Therefore, for any number of mutations, privacy loss is determined entirely by the sampled-Gaussian score releases.

\vspace{-0.3em}
\subsection{Proof of Theorem~\ref{thm:main}}
Mutations consume zero budget (Proposition~\ref{prop:mutation-free}). Each executed iteration releases $K$ sampled-Gaussian scores. Applying the without-replacement RDP amplification bound to each release, composing the resulting RDP guarantees over at most $TK$ adaptive calls, and converting at the target $\delta$ yields Table~\ref{tab:rdp_bounds}. Selection and subsequent mutation are post-processing, so the final prompt inherits the same $(\varepsilon,\delta)$ guarantee.

\end{document}